\documentclass[12pt,epsf]{article}
\usepackage{amssymb}
\usepackage{graphicx}
\addtocounter{page}{0} \tolerance=5000
\newcommand{\beq}{\begin{equation}}
\newcommand{\eeq}{\end{equation}}
\newcommand{\beqn}{\begin{eqnarray}}
\newcommand{\eeqn}{\end{eqnarray}}
\newcommand{\x}{(p_\mu p_\pi)}
\newcommand{\y}{(p_k p_\mu)}
\newcommand{\z}{(p_k p_\pi)}
\newcommand{\PP}{P^2}
\newcommand{\Ps}{ P }
\newcommand{\mm}{m_\mu^2}
\newcommand{\mq}{m_\pi^2}
\newcommand{\mk}{m_k^2}
\newcommand{\I}{I_{\mu\pi}}

\newcommand{\CC}{\frac{C}{D}}

\newcommand{\Bb}{\frac{1}{I^2}\left(1-\frac{\mm}{\PP}\right)\left((p_k P)-\frac{1}{2}\mk\PP L_4\right)}

\newcommand{\al}{\mbox{${\alpha}$}}

\title{ Transverse muon polarization in
$K^{\pm}\to\pi^0\mu^{\pm}\nu$ decay induced by the two-photon
final-state interaction}
\author{V.P.~Efrosinin$^a$, I.B.~Khriplovich$^{b,c}$, G.G.~Kirilin$^{b,c}$, Yu.G.~Kudenko$^a$}
\date{}
\begin{document}
\maketitle
\begin{center}
$^a$Institute for Nuclear Research, 117312 Moscow, Russia\\
$^b$Budker Institute of Nuclear Physics, 630090 Novosibirsk,
Russia\\
$^c$Novosibirsk University, 630090 Novosibirsk, Russia
\end{center}
\begin{abstract}
We calculate the transverse muon polarization in the decay
$K^{\pm}\to\pi^0\mu^{\pm}\nu$ induced by the electromagnetic
two-photon final-state interaction.   For the central part of the Dalitz plot
the typical value of this polarization is about $4\times 10^{-6}$.
\end{abstract}
\section{Introduction}

Measurement of the muon transverse polarization $P_T$ in the decay
$K^{\pm}\to\pi^0\mu^{\pm}\nu$~($K_{\mu3}$) can provide important
insight to new physics beyond the Standard Model (SM). In the case of
$K_{\mu3}$ decay, $P_T$ is a T-odd observable ${\bf s}_{\mu}\cdot
({\bf p}_{\pi}\times{\bf p}_{\mu})$ determined by the momenta ${\bf
p}_{\pi}$ of the $\pi^0$, ${\bf p}_{\mu}$ of the $\mu$  and the spin ${\bf
s}_{\mu}$ of the $\mu$. This observable is very small in the SM, but
it is an interesting probe of non-SM CP-violation
mechanisms~\cite{nonsm} where $P_T$  can be as large as $10^{-3}$
in either $K_{\mu3}$ or ~$K^{\pm}\to
\mu^{\pm}\nu\gamma$~($K_{\mu2\gamma}$). The recent measurement of $P_T$ in the
$K^+\to\pi^0\mu^+\nu$ decay provided by KEK E246~\cite{e246}
gives a value of $P_T$ consistent with zero with
experimental error  $\sigma\sim 5\times 10^{-3}$ and will be
further improved by a factor of 4. The first limit on $P_T$ in
$K_{\mu2\gamma}$  will   also be obtained in this experiment.
Proposed experiments~\cite{kud} could reach sensitivity to
$P_T$ of $\leq 10^{-4}$.

\begin{figure}[htpb]
\centering\includegraphics[width=10cm,angle=270]{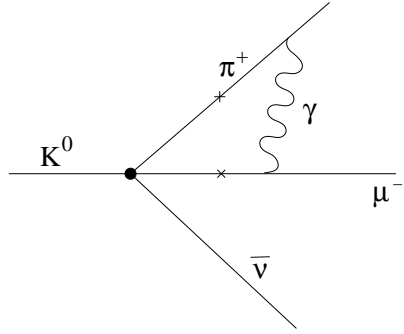}
\caption{Final-state electromagnetic interaction producing
$P^{em}_T$ in the
decay $K^0\to\pi^+\mu^-\bar\nu$. Crosses mark on-mass-shell particles.}
\label{ko}
\end{figure}

\begin{figure}[htpb]
\centering\includegraphics[width=10cm,angle=270]{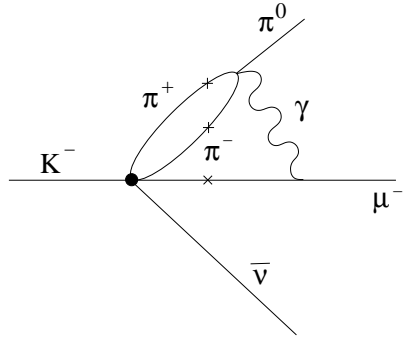}
\caption{The contribution of the two-loop diagram to $P^{em}_T$ in
$K^-\to\pi^0\mu^-\bar\nu$ decay. Crosses mark
on-mass-shell particles.}
\label{kpipi}
\end{figure}

In fact, whether CP or T is violated or not, a nonvanishing $P_T$
in $K_{\mu3}$ decays can be induced by electromagnetic final-state
interactions (FSI). In the neutral kaon decays this correlation
arises due to the imaginary part of the diagram in Fig.~\ref{ko}, and can
be as large as $\sim 10^{-3}$~\cite{by,okun}. In this sense,
$K^{\pm}\to\pi^0\mu^{\pm}\nu$ decays have a major advantage:
here the final pion is neutral, and thus there is no elastic
electromagnetic FSI. The single-photon contribution to $P^{em}_T$
arises in these decays due to the imaginary part of the two-loop
diagram shown in Fig.~\ref{kpipi}, and was estimated as $P^{em}_T \le
10^{-6}$~\cite{zhi}. Such a strong suppression of this imaginary
part is caused by the small value of the phase space of the
three-particle intermediate state.

In the present work we demonstrate that the two-photon effects
result in a larger contribution to the transverse muon
polarization in the $K^{\pm}\to\pi^0\mu^{\pm}\nu$ decays.

Let us note that the FSI contribution to the muon transverse
polarization in the $K^+\to\pi^0\mu^+\nu$ and
$K^-\to\pi^0\mu^-\bar{\nu}$ decays is the same: $P^{em+}_T=
P^{em-}_T$, while the CP-odd contributions are of opposite signs:
$P^{CP+}_T=- P^{CP-}_T$. To be definite, we present all
intermediate formulae for the $K^- \to \pi^0 \mu^- \bar\nu$ decay.

For the $K_{\mu3}$ decay, the most general invariant amplitude is
\beq
\label{kmu3ampl/1} M_{K_{\mu3}} =  -\frac{G_F}{\sqrt2}V_{us}
 \frac{1}{\sqrt2}\Bigl[(p_K +p_{\pi} )_{\mu} f_+(t)+(p_K
-p_{\pi})_{\mu} f_-(t) \Bigr]\,\bar{u}(p_{\mu}) \gamma_{\mu}
(1+\gamma_5 )v(p_{\nu}).
\eeq
Here $G_F$ is the Fermi constant; $V_{us}$ is the element of the
Cabibbo-Kobayashi-Maskawa matrix; $f_+(t)$ and  $f_-(t)$ are form
factors,
 $p_K,\; p_{\pi},\; p_{\mu},$ and
$p_{\nu}$ are the momenta of the kaon, pion, muon, and
antineutrino, respectively; $t=(p_K -p_{\pi})^2$ is the momentum
transfer squared to the lepton pair. Our convention for
$\gamma_5$ is
\beq\label{g5}
\gamma_5 = \left(\begin{array}{rr}0  & -I \\
 -I &  0 \end{array}\right).
\eeq
Expression (\ref{kmu3ampl/1}) can be conveniently rewritten as
\begin{equation}
\label{kmu3ampl/2} M = -\,{G_F \over \sqrt{2}}\,V_{us}
\sqrt{2}\,f_+(t) \bar u (p_{\mu}) (\hat p_K +\chi \hat
p_{\mu})(1+\gamma_5)v(p_{\nu}),
\end{equation}
where
\beq\label{xi}
\chi =\,\frac{1}{2} (\xi -1), \quad \xi =\,\frac{f_-}{f_+}
=-0.35\pm 0.15~\cite{part}.
\eeq
The standard parameterization is
\begin{equation}
f_+(t) = f_+(0)\left(1 + \lambda \frac{t}{m^2_{\pi}}\right), \quad
\lambda =0.0286 \pm 0.0022~\cite{part}.
\end{equation}
Experimental data are compatible with a constant, i.e.,
$t$-independent, $f_-\,$.

The covariant, 4-dimensional form of the transverse polarization
vector is
\beq P_{T\alpha}=- 2\mathrm{Im}\,\chi m_\mu
\varepsilon_{\alpha\beta\gamma\delta} p_{\mu\beta} p_{\nu\gamma}
p_{k\delta}/\Phi,
\end{equation}
\[
\varepsilon_{0123}=1, \quad \Phi =  2(p_{\mu} p_K)(p_{\nu}
p_K)-m_K^2 (p_{\mu} p_{\nu})+2\chi m_{\mu}^2 (p_{\nu}
p_K)+|\chi|^2 m_{\mu}^2 (p_{\mu}p_{\nu}).
\]
The covariant expression for the degree of transverse
polarization is
\beq
P_T =2\mathrm{Im}\chi m_\mu \sqrt{D}/\Phi\,,
\eeq
where
\beq\label{D}
D= m_{\mu}^2 m^2_{\pi} m_K^2 + 2(p_{\mu} p_{\pi})(p_{\mu}
p_K)(p_{\pi}p_K)-m_{\mu}^2(p_{\pi}p_K)^2 - m^2_{\pi}(p_{\mu}
p_K)^2 - m_K^2(p_{\mu} p_{\pi})^2 . \eeq

\section{Some details of the calculation}
\begin{figure}[htpb]
\centering\includegraphics[width=14cm,angle=0]{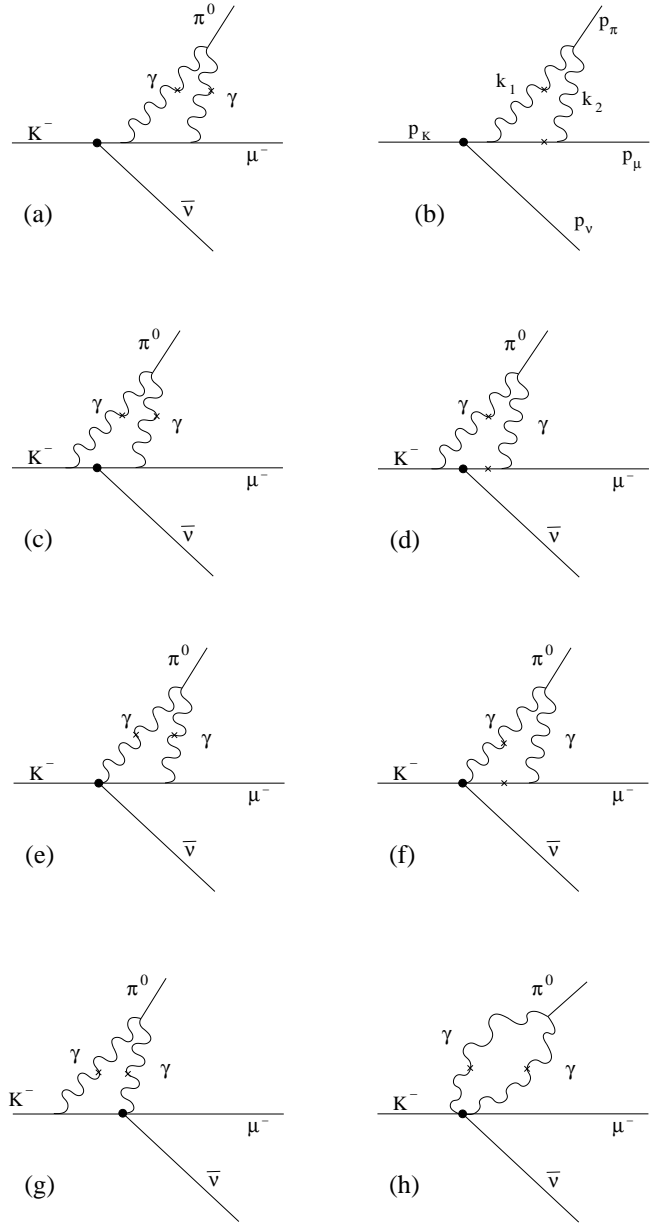}
\caption{The contribution of two-photon final-state interactions to
$P^{em}_T$ in $K^-\to\pi^0\mu^-\bar\nu$ decay. Crosses mark
on-mass-shell particles. }
\label{photon}
\end{figure}

We are interested in the transverse polarization due to
$\mathrm{Im}\chi$ induced by the FSI with two photons. Here the
transverse muon polarization is proportional to the imaginary
parts of diagrams shown in Fig.~\ref{photon}. In diagrams a,~c,~e,~g,~h
both
photons are real and the intermediate muon is off-mass-shell. In
diagrams b,~d,~f the intermediate muon and one photon are
on-mass-shell, and the second photon is virtual. In the diagrams,
the on-mass-shell intermediate particles are marked by crosses.
Among the diagrams, those with both photons attached to the kaon,
are absent. Obviously, there is no P-even effective $KK\pi$
vertex.

The common vertex for the diagrams~\ref{photon}a,~b is the
$\pi^0\rightarrow\gamma\gamma$ amplitude:
\begin{equation}\label{pi2g}
M(\pi^0\rightarrow\gamma\gamma)=-\sqrt{2}\,{\alpha \over \pi
f_\pi}\,\varepsilon_{\alpha\beta\gamma\delta}
\,e_{1\alpha}\,e_{2\beta}\,k_{1\gamma}\,k_{2\delta},
\end{equation}
where $e_{1\alpha},\;e_{2\beta}$ are the polarization vectors of
the photons, $k_{1\gamma},\;k_{2\delta}$ are their momenta and
$f_{\pi} = 0.13$~GeV. We use here the theoretical expression for
the coupling constant; the $\pi_0$ life time calculated with it
reproduces the experimental value within the accuracy of the
latter.

The $K^- \to \mu^- \nu$ vertex in the Inner Bremsstrahlung (IB)
diagrams~\ref{photon}a-d, is
\beq
M(K_{\mu2})=-\,\frac{G_F}{\sqrt2}\,V_{us}if_K p_{K\mu}
\bar{u}(p_{\mu}) \gamma_{\mu} (1+\gamma_5 )v(p_{\nu}),
\eeq
where $f_K=0.16$ GeV.

The $K^- \to \mu^- \bar{\nu} \gamma$ vertex in the structure
dependent ($SD$) radiation diagrams in Fig.~\ref{photon}e-g is described by two
independent amplitudes:
\beq\label{SDV}
M(SD_V)=-{G \over \sqrt{2}}V_{us}\sqrt{4\pi\al}F_V
\varepsilon_{\alpha\beta\gamma\delta}
\,e_{1\alpha}\,\bar{u}(p_{\mu}) \gamma_{\beta} (1+\gamma_5)
v(p_{\nu}) \,k_{1\gamma}\,(p_K-k_{1})_{\delta};
\eeq
\beq\label{SDA}
M(SD_A)=-{G \over \sqrt{2}}V_{us}\sqrt{4\pi\al}(-i)F_A
[\delta_{\alpha\beta}\,(p_K k_1)- (p_K-k_{1})_{\alpha}k_{1\beta}]
\,e_{1\alpha}\,\bar{u}(p_{\mu}) \gamma_{\beta} (1+\gamma_5)
v(p_{\nu}).
\eeq
Let us note here that when going over from the $K^- \to \mu^-
\bar{\nu} \gamma$ decay to $K^+ \to \mu^+  \nu \gamma$, not
only the lepton current in both amplitudes changes to
$\bar{u}(p_{\nu}) \gamma_{\beta} (1+\gamma_5) v(p_{\mu})$, but in
the second amplitude, $M(SD_A)$, $(-i)$ changes to $i$.
Theoretically, the vector form factor $F_V$ is directly related
to the $\pi^0\rightarrow\gamma\gamma$ amplitude~\cite{vi} (see
also~\cite{mu,baen}). In the case of $\pi \to e \nu \gamma$
decay one should delete  $4\pi\al$ from the constant
$\sqrt{2}\al/\pi f_\pi$ in~\cite{vi}, and then divide by
$\sqrt{2}$. In our case of the $K \to \mu \nu \gamma$ decay one
should also change $f_\pi \to f_K$. Thus we obtain
\beq
F_V=\,{1 \over 4\pi^2 f_K}\,=\,{0.079 \over m_K}.
\eeq
The value of the axial form factor $F_A$, as predicted by
chiral perturbation theory, is~\cite{bijn}
\begin{equation}\label{chfa}
F_A=\frac{0.043}{m_K}.
\end{equation}
The experimental study of the decay $K^+\to
\mu^+\nu\gamma$~\cite{e787} results in the following information on
the corresponding form factors in this decay:
$$-0.04/m_K<F_V-F_A<0.24/m_K, \quad |F_V+F_A|=(0.165\pm 0.007\pm
0.011)/m_K.$$

As to the diagram~\ref{photon}h with a double SD radiation,
qualitative arguments will be presented below which allow one to
neglect it.

A common feature of all diagrams where both intermediate photons
are real (\ref{photon}a,~c,~e,~g,~h), is that their contributions are
proportional to $m_\pi^2$. Indeed, in the limit $m_\pi \to 0$ the
4-momenta of the photons in the $\pi^0 \to \gamma \gamma$ decay
become collinear, and the matrix element (\ref{pi2g}) vanishes.
On the scale of the $K_{\mu 3}$ problem, $m_\pi^2$ is relatively
small, and correspondingly, the contributions of
 diagrams~\ref{photon}a,~c,~e,~g,~hare rather small as well.

On the other hand, there is a certain similarity between
contributions of the intermediate states with two real photons
and real photon and muon in diagrams of the same structure, i.e.,
between imaginary parts corresponding to  diagrams of Fig.~\ref{photon}:
 a and b, c and d,
e and f, respectively. Therefore, it is expedient to calculate the
respective pairs of imaginary parts in parallel. Strong
cancellations occur between the terms proportional to $m_\pi^2$
in diagrams a and b, c and d, e and f. In particular, the most formidable
integrals, which arise in diagrams a and b, cancel completely.

Still, the calculation remains quite tedious, and only its final
results will be presented.

\section{Internal Bremsstrahlung}
We start with the contribution of the IB diagrams~\ref{photon}a-d:
\begin{eqnarray}\label{chib}
2\,\mathrm{Im}\,\chi(IB) & = & -\frac{\alpha^2}{4\pi}\,{f_K \over
f_\pi}\,{1 \over f_+(t)} \Bigl\{4+3\chi\nonumber\\
& & +\frac{1}{\I^2}\Big[\x\mq+\chi\mm\mq-\frac{\x\mm\chi(p_\pi
P)}{\PP}\Big]-\frac{2\mm}{\PP}\nonumber\\
& & +\frac{\mq
L_2}{2}\Bigl[1-\frac{\mq}{\I^2}\left(\x+(1+\chi)\mm\right)
+\frac{2(1+\chi)\mm}{\mq+2\x}\Bigr]\nonumber\\
& & +\frac{2\x\z-\mq\y}{2D}\,\Bigl[L_2\Bigl(\z\x-\mq\y+\chi\,\I^2\nonumber\\
& & -(1+\chi)\bigl(\y\x-\mm\z\bigr)\Bigr)+L_3\Bigl(\chi\bigl(\y\mq-\x\z\bigr)\nonumber\\
& & -I_{k\pi}^2-(1+\chi)\bigl(\y\z-\mk\x\bigr)\Bigr)\nonumber\\
 & & +L_4\Bigl(I^2+\chi\bigl[\bigl(\z-\mk)(p_\mu P)-\y((p_\pi P)-(p_k
P)\bigr)\bigr]\Bigr)\Bigr]\nonumber\\ & & +[\z(1+\chi)-\frac{\mq
\chi}{2}](L_3-L_4)-L_4\left[\y(1+\chi)-\chi\x\right]\nonumber\\ &
&+\frac{\chi}{I^2}[\x-\y+\mm]
\left(1-\frac{\mm}{\PP}\right)\left[(p_k\Ps) -\frac{1}{2}\mk\PP
L_4\right]\nonumber\\ & &
+\left[\Bb-\CC\right]\left[\chi\bigl(\z-\x-\mq\bigr)\right.\nonumber\\
& &\left.- (1+\chi) \bigl(\mk-(p_k P)\bigr)\right]\Bigr\}.
\end{eqnarray}
Here $\chi=\chi(t)$ is defined in fact by the same relation
(\ref{xi}) and does not include $\mathrm{Im}\,\chi$ which is being
calculated, $D$ is given in (\ref{D}), $P= p_\mu + p_\pi$,
\[
L_2=\frac{1}{\I}\ln\frac{(p_\pi P)+\I}{(p_\pi P)-\I}\,, \quad
L_3=\frac{1}{I_{ k \pi}}\ln\frac{\z+I_{k\pi}}{\z-I_{k\pi}}\,,
\quad L_4=\frac{1}{I}\ln\frac{(p_k P)+I}{(p_k P)-I}\,, \]
\[\I=\sqrt{\x^2-\mm\mq}\,, \quad I_{k\pi}=\sqrt{\z^2-\mk\mq}\,,
\quad  I=\sqrt{(p_k P)^2-P^2\mk}\,,\]
\begin{eqnarray*}
C & = &
\frac{1}{2}\left\{L_2\left[\x\y\mq-2\x^2\z+\z\mq\mm\right]\right.\nonumber\\
& &
-\left(L_3-L_4\right)\left[\z\y\mq-2\x\z^2+\x\mq\mk\right]\nonumber\\
& & +L_4\left.\left[\x^2\mk-\mm\z^2-D\right]\right\}\,.
\end{eqnarray*}

The first line $4+3\chi$ in braces in (\ref{chib}) is dominating,
it constitutes about 70\% at the center of the Dalitz plot. The
typical value of the IB contribution in this region is
\beq
2\,\mathrm{Im}\,\chi(IB)=1.4\times 10^{-5}.
\eeq

\section{Structure-dependent radiation}
The vector SD amplitude (\ref{SDV}) is operative in all three
diagrams~\ref{photon}e-g. Its contribution equals
\begin{eqnarray}\label{chSDV}
2\,\mathrm{Im}\,\chi(SD_V) & = & \frac{\alpha^2 F_V}{4\pi f_\pi
f_+(t) }\,
\Biggl\{\frac{1}{2}\left[\left(1-\frac{\mm}{\PP}\right)^2-{\mq
\over \I^2}\,\left(\frac{\x\mq L_2}{2}-\x\right.\right.\nonumber\\
& & \left.\left.+\frac{\mm(p_\pi P)}{\PP}\right)\right](\PP+\chi(P
p_\mu))+\frac{\chi L_2 m_\pi^4}{4}\nonumber\\ & &
+\frac{\chi}{2}\,\Bigl(1-\frac{\mm}{\PP}\Bigr) \Biggl[\mq +
\Bigl(1-\frac{\mm}{\PP}\Bigr)\frac{(P p_\pi)}{2}
+\frac{1}{3}\Bigl(1-\frac{\mm}{\PP}\Bigr)^2\nonumber\\ & &
\times\Bigl(2\mk+\PP\Bigr)\Biggr]+A\left[(P p_k)+\chi\y\right]
+\,{m_\pi^2 \over I_{k\pi}^2}\left[(p_k p_\pi)-\,{m_k^2 m_\pi^2
\over 2}\,L_3\right] \nonumber\\ & & \times \left[\chi\left((p_k
p_\pi)- m_\pi^2\right) + (1+\chi)\left((p_k p_\pi)-
m_k^2\right)\right] \Biggr\}\,.
\end{eqnarray}
Numerically, this contribution at the center of the Dalitz plot is
\beq
2\,\mathrm{Im}\,\chi(SD_V)=-0.9\times 10^{-6}.
\eeq

Let us consider now the contribution of the axial SD amplitude
(\ref{SDA}). In this case diagram~\ref{photon}g is not operative, and
diagrams~\ref{photon}e,f give
\begin{eqnarray}\label{chSDA}
2\,\mathrm{Im}\,\chi(SD_A) & = & \frac{\alpha^2 F_V}{4\pi f_\pi
f_+(t) }\,
\Biggl\{\frac{1}{2}\left[\left(1-\frac{\mm}{\PP}\right)^2-{\mq
\over I_{pq}^2}\,\left(\frac{\x\mq L_2}{2}-\x\right.\right.\nonumber\\
& & \left.\left.+\frac{\mm(\mq+\x)}{\PP}\right)\right](\PP+\chi(P
p_\mu))+\frac{\chi L_2 m_\pi^4}{4}\nonumber\\ & &
+\frac{\chi}{2}\,\Bigl(1-\frac{\mm}{\PP}\Bigr) \Biggl[\mq +
\Bigl(1-\frac{\mm}{\PP}\Bigr)\frac{(P p_\pi)}{2}
+\frac{1}{3}\Bigl(1-\frac{\mm}{\PP}\Bigr)^2
\Bigl(2\mk+\PP\Bigr)\Biggr]\nonumber\\
& &+A\left[(P p_k)-\mk-\chi\left(\y-(P p_\mu)\right)\right]
\Biggr\}\,.
\end{eqnarray}
Here
\begin{eqnarray}
A & = & \frac{1}{\I^2}\Biggl\{2 \x^2 -\mm(P
p_\pi)\left(1-\frac{\mm}{\PP}\right)\Biggr\}-{\mm m_\pi^4 L_2
\over 2 I_{pq}^2}\,.
\end{eqnarray}
If one assumes for $F_A$ its theoretical value (\ref{chfa}), then
the contribution of (\ref{chSDA}) constitutes numerically at the
center of the Dalitz plot
\beq
2\,\mathrm{Im}\,\chi(SD_A) \simeq 0.3\times 10^{-6}.
\eeq

And at last, the double structure-dependent (DSD) radiation (see
diagram~\ref{photon}h). The small magnitude even of the single SD effects, gives
us serious reasons to assume that the contribution of the DSD
diagram~\ref{photon}h to $2\mathrm{Im}\chi$ can be safely neglected, so much
the more that it is proportional to $m_\pi^2$.


\section{Conclusion}
Among the effects of the FSI, it is the two-photon ones which
provide the main contribution to the transverse muon polarization
in $K^{\pm}\to\pi^0\mu^{\pm}\nu$ decay. The typical value of
$2\,\mathrm{Im}\chi$ in the central part of the Dalitz plot is
close to
\beq
2\,\mathrm{Im}\chi=2\,\mathrm{Im}\chi(IB)+2\,\mathrm{Im}\chi(SD_V)+
2\,\mathrm{Im}\chi(SD_A) \simeq 1.3\times 10^{-5}.
\eeq
The deviation from this value over the Dalitz plot does not exceed
15\%.
\begin{figure}[ht]
\centering\includegraphics[width=12cm,angle=0]{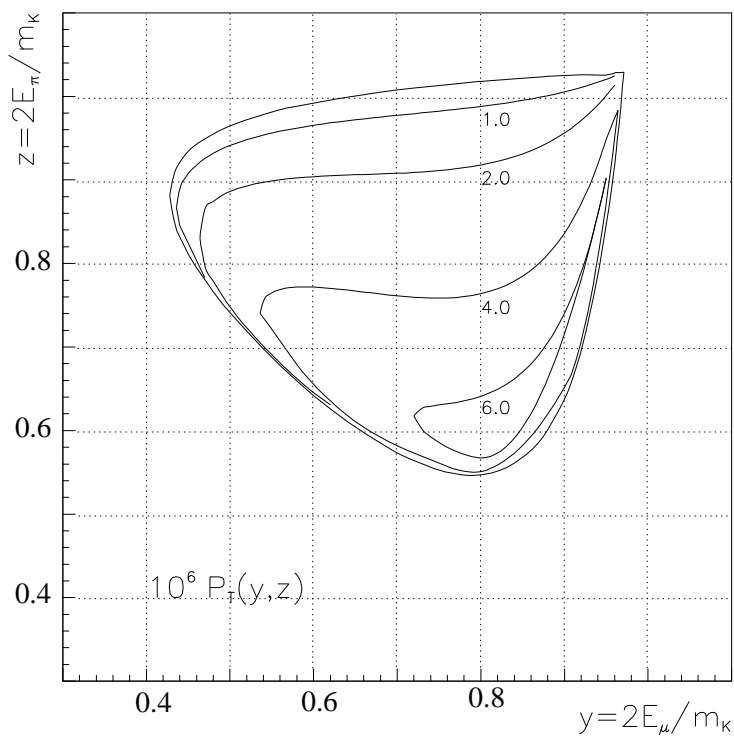}
\caption{ Dalitz plot for $P^{em}_T$ in
$K^{\pm}\to\pi^0\mu^{\pm}\nu$ decay. Contours correspond to
constant $P^{em}_T$.} \label{dalitz}
\end{figure}
~\\
~\\
The typical value of the transverse muon polarization in the
central part of the Dalitz plot is about
\beq
P^{em}_T \simeq 4\times 10^{-6}.
\eeq
The distribution of this polarization over the Dalitz plot is
shown in Fig.~\ref{dalitz}. At low pion energy it can reach the level of
$6\times 10^{-6}$.

\begin{center}***\end{center}
This work was supported in part by the Russian Foundation for
Basic Research through Grants No.~98-02-17797 and 99-02-17814,
I.B.K. and G.G.K. acknowledge also the support by the Ministry of
Education through Grant No.~3N-224-98, and by the Federal Program
Integration-1998 through Project No.~274.

\section*{Appendix}
Here we present for convenience some useful integrals.

\bigskip

\noindent 1. Two-photon intermediate states

\[  d\rho_{(\gamma\gamma)}=\frac{1}{(2\pi)^2}\,\delta(p_\pi-k_1-k_2)
\frac{d^3k_1}{2\omega_1}\frac{d^3k_2}{2\omega_2}\,,\quad
k_1^2=0,\quad k_2^2=0.  \]
\[ \int d\rho_{(\gamma\gamma)}=\,{1 \over 8\pi}\,. \]
\[ \int k_{1\alpha}\,d\rho_{(\gamma\gamma)}
=\frac{p_{\pi\alpha}}{16 \pi}\,.\]
\[ \int\frac{d\rho_{(\gamma\gamma)}}{(p_\mu
k_2)} =\frac{1}{8\pi}L_1.\]
\[ \int\frac{k_{2\alpha}}{(p_\mu
k_2)}\,d\rho_{(\gamma\gamma)} =\,\frac{p_{\mu\alpha}}{8\pi
\I^2}\Biggl[\frac{\mq\x}{2}L_1-\mq
\Biggr]+\frac{p_{\pi\alpha}}{8\pi
\I^2}\Biggl[\x-\frac{\mm\x}{2}L_1 \Biggr].\]
\[ \int\frac{d\rho_{(\gamma\gamma)}}{(p_k k_1)}=\frac{1}{8\pi}L_3.\]
\[\int\frac{k_{1\alpha}}{(p_k k_1)}\,d\rho_{(\gamma\gamma)}
=\,\frac{p_{k\alpha}}{8\pi
I^2_{k\pi}}\Biggl[\frac{\mq\z}{2}L_3-\mq \Biggr]
+\frac{p_{\pi\alpha}}{8\pi I^2_{k\pi}}\Biggl[\z-\frac{\mk\x}{2}L_3
\Biggr]. \] Here $$L_1=\frac{1}{\I}\ln\frac{\x+\I}{\x-\I}\,.$$
$\I$, $I_{k\pi}$, and $L_3$ are defined in the text.

\bigskip

\noindent 2. Muon-photon intermediate states

\bigskip

\[ d\rho_{(\mu\gamma)}=\frac{1}{(2\pi)^2}\,\delta(p_\mu+p_\pi-k_1-p)
\frac{d^3k_1}{2\omega_1}\frac{d^3p}{2p_0}\,,\quad
p^2=(p_\mu+k_2)^2=\mm, \quad k_1^2=0. \]
\[ \int d\rho_{(\mu\gamma)}\,=\,\frac{1}{8\pi}\,\left(1-\frac{\mm}{\PP}\right)\,.\]
\[ \int k_{1\alpha}\,d\rho_{(\mu\gamma)}=\,{1 \over 16\pi}
\left(1-\frac{\mm}{\PP}\right)^2 P_\alpha . \]
\[ \int k_{1\alpha}k_{1\beta}\,d\rho_{(\mu\gamma)}=\,{1 \over 96\pi}\,
\left(1-\frac{\mm}{\PP}\right)^3 \left(g_{\alpha\beta}\PP - 4
P_\alpha P_\beta\right). \]

\[\int\frac{d\rho_{(\mu\gamma)}}{(p_\pi-k_1)^2}=-\frac{1}{16\pi}\,(L_1+L_2).\]
\begin{eqnarray*}
\int\frac{k_{2\alpha}}{(p_\pi-k_1)^2}\,d\rho_{(\mu\gamma)} & = &
\,{1 \over 16 \pi \I^2}\left\{
p_{\mu\alpha}\Biggl[\left(1-\frac{\mm}{\PP}\right) (p_\pi P)
-\,{\x\mq \over 2}\,(L_1+L_2)\Biggr] \right.\nonumber\\ & &
\left.-p_{\pi\alpha}\Biggl[\left(1-\frac{\mm}{\PP}\right)(p_\mu P
)-\,{\mm\mq \over 2}\,(L_1+L_2)\Biggr]\right\}. \end{eqnarray*}
\[ \int\frac{d\rho_{(\mu\gamma)}}{(p_k k_1)}=\frac{1}{8\pi}L_4.\]
\[ \int\frac{k_{1\alpha}}{(p_k k_1)}\,d\rho_{(\mu\gamma)}
={1 \over 8\pi I^2}\,\left(1-\frac{\mm}{\PP}\right)\,
\left\{p_{k\alpha}\PP \left({1 \over 2}\,(p_k P)L_4-
1\right)\right.\]
\[\left. + P_{\alpha}\left[(p_k P)-\,{1 \over 2}\,m_k^2\PP L_4
\right]\right\}. \]

\bigskip

\noindent 3. Integrals generating constants $C$ and $A$

\bigskip

\[\int \frac{\varepsilon_{\alpha\beta\gamma\delta}p_{k\alpha}
p_{\mu\beta} p_{\pi\gamma} k_{1\delta}
k_{1\rho}d\rho_{(\gamma\gamma)}
-\varepsilon_{\alpha\beta\gamma\delta} p_{k\alpha} p_{\mu\beta}
p_{\pi\gamma} k_{1\delta} k_{1\rho} d \rho_{(\mu\gamma)}}{(p_k
k_1)(p_\mu k_2)}= - \frac{C}{8\pi
D}\,\varepsilon_{\alpha\beta\gamma\rho}p_{k\alpha} p_{\mu\beta}
p_{\pi\gamma},.\]
\[ \int
\frac{\varepsilon_{\alpha\beta\gamma\delta}p_{k\alpha}
p_{\mu\beta} p_{\pi\gamma} k_{1\delta} k_{1\rho} d
\rho_{(\gamma\gamma)} -\varepsilon_{\alpha\beta\gamma\delta}
p_{k\alpha} p_{\mu\beta} p_{\pi\gamma} k_{1\delta} k_{1\rho} d
\rho_{(\mu\gamma)}}{(p_\mu k_2)}=
\frac{A}{32\pi}\,\varepsilon_{\alpha\beta\gamma\rho}p_{k\alpha}
p_{\mu\beta} p_{\pi\gamma}\,. \]

Explicit expressions for $C$, $D$, and $A$ are given in the text.

\end{document}